# Kinetics of phase transition from lamellar to hexagonally packed cylinders for a triblock copolymer in a selective solvent

Yongsheng Liu, Minghai Li, Rama Bansil[*]

*Department of Physics, Boston University, Boston, MA 02215, USA*

Milos Steinhart

*Institute of Macromolecular Chemistry, Academy of Sciences of the Czech Republic, Heyrovsky Sq. 2, 162 06 Prague 6, Czech Republic*

**Abstract:** We examined the kinetics of the transformation from the lamellar (LAM) to the hexagonally packed cylinder (HEX) phase for the triblock copolymer, polystyrene-*b*-poly (ethylene-*co*-butylene)-*b*-polystyrene (SEBS) in dibutyl phthalate (DBP), a selective solvent for polystyrene (PS), using time-resolved small angle x-ray scattering (SAXS). We observe the HEX phase with the EB block in the cores at a lower temperature than the LAM phase due to the solvent selectivity of DBP for the PS block. Analysis of the SAXS data for a deep temperature quench well below the LAM-HEX transition shows that the transformation occurs in a one-step process. We calculate the scattering using a geometric model of rippled layers with adjacent layers totally out of phase during the transformation. The agreement of the calculations with the data further supports the continuous transformation mechanism from the LAM to HEX for a deep quench. In contrast, for a shallow quench close to the OOT we find agreement with a two-step nucleation and growth mechanism.

---

[*] Author to whom correspondence should be addressed. Email: rb@bu.edu

**Introduction**

Block copolymers exhibit well known microdomain structures,[1-4] including classic equilibrium phases of lamellar (LAM), hexagonally packed cylinder (HEX), body-centered-cubic and face-centered-cubic, and more unusual mesophases of gyroid and hexagonally perforated lamellar (HPL). The morphology of block copolymer depends on the composition, the incompatibility of the two blocks, molecular weight, and temperature.[1-4] In the case of block copolymers in selective solvents, the morphology also depends on the volume concentration and the selectivity of the solvents.

There have been many studies on the phase behavior of block copolymers.[2,3,5,6] Phase transition between the LAM and HEX phases of block copolymer melts has been extensively studied using SAXS and transmission electron microscopy (TEM) techniques.[7-16] The transition from the LAM and HEX phases to an intermediate perforated HPL mesophase has also been examined using TEM and small angle x-ray scattering (SAXS).[17,18] Most of these studies were dealing with diblock copolymer melts and show that block copolymer melts form the LAM phase at lower temperature and transform into the HEX phase upon heating.

Theoretical models have been developed by several groups to explain the mechanism of the transformation from the LAM to the HEX phase.[19-24] A commonly accepted model is that the lamellae in the LAM phase break into *in-plane* cylinders and transform into the HEX phase. There are different opinions on how lamellae break into cylinders. Hamley et al.[8] used Fourier synthesis to reconstruct the transformation from scattering images and concluded that the cylinders in adjacent lamellae are correlated and out of phase; Qi and Wang[20, 21] and Luo and Yang[24] utilized time dependent Ginzburg-Landau (TDGL) approach to simulate the LAM to the HEX phase transition and stated that the adjacent lamellae are correlated during the transition. In this way, the transformation of the LAM to the HEX is a single step process. On the other hand, Hajduk et al.[10] employed TEM technique to visualize the LAM to the HEX transition and saw that lamellae in the LAM phase break into cylinders randomly and rearrange into the HEX structure. Using a model with anisotropic fluctuations, Laradji et al.[22] simulated this transition and concluded that adjacent lamellae are not correlated. If this holds true, the LAM to the HEX transformation is a two step process. Although it is a simple geometric model, surprisingly, no one has tried to calculate the scattering of the model to validate these results from SAXS data.

To the best of our knowledge, there is no study of the kinetics of the LAM to HEX transformation in a block copolymer in a selective solvent. Previous studies of kinetics in melts have been reported.[15,19] In this paper, we present a time-resolved SAXS study of the LAM to HEX transformation in the triblock copolymer of (polystyrene-*b*-poly(ethylene-*co*-butylene)-*b*-polystyrene) (SEBS) in dibutyl phthalate (DBP) which is a good solvent for the outer polystyrene (PS) blocks. Unlike the block copolymer melts this solvent selective system exhibits a HEX phase at lower temperature than a LAM phase. Thus the system transforms from the LAM phase to the HEX phase upon cooling. The slower rate of transformation in a cooling jump made it feasible to examine the kinetics of the LAM to HEX transition by using time-resolved SAXS measurements. We compare the results with a calculation of the scattering using a geometrical model.

**Experimental Section**

**Materials.** The triblock copolymer of SEBS (Shell chemicals, Kraton G1650 with a molecular weight $M_n$ 100 kg/mol, polydispersity $M_w/M_n$ = 1.05, PS fraction 28 wt%, E:B ratio 1:1) 40% w/v, was cast into DBP which is selective to outer PS blocks. Methylene chloride (dichloromethane, 99.6%) was added into the solution to dissolve the polymer. An antioxidant 2,6-di-tert-butyl-4-methyphenol (BHT), 0.5% wt, was also added to prevent oxidization of the polymer. The mixture was heated at 100 $^o$C for several hours and shaken till a clear solution was obtained. Then the solution was evaporated to remove methylene chloride, until there was no weight change of the sample over at least 24 hours.

**Small Angle X-ray Scattering.** Time-resolved SAXS experiments were carried out at X27C beamline of the National Synchrotron Light Source (NSLS) at Brookhaven National Laboratory. The X-ray wavelength was 0.1366 nm (9.01 keV) with energy resolution of 1.1%. A two dimensional MAR CCD detector with an array of 1024 x 1024 was used to record the scattering pattern. Azimuthal integration of the image was performed to obtain the scattering intensity I(q) versus the scattering vector $q = (4\pi/\lambda)sin\theta$, with 2θ being the scattering angle. In our experimental setup with a sample-to-detector distance of 1.957 m we cover a q-range of 2 nm$^{-1}$ with q=0 at the center of the 1024 X 1024 pixel CCD detector. This sets the limiting q-resolution Δq per pixel = 0.004 nm$^{-1}$. Details of the experimental apparatus including the sample cell, the temperature control system and data analysis are described in previous works from our laboratory.[25,26]

Measurements at fixed value of temperature were used to identify the phases of the polymer solution. The time evolution of SAXS data was measured following either a temperature ramp (T-ramp) or a temperature jump (T-jump) protocol. T-ramp SAXS experiments were used to determine the order-order transition (OOT) temperature. T-jump experiments crossing the transition temperature were done to observe the kinetics of the OOT process. In both measurements we recorded the scattering pattern for 10 seconds per frame; data transfer from the 2D MAR CCD array to PC took an additional 7 s/frame. The total duration of each run was about 1.5 to 2 hours.

**Data processing and fitting.** We used Matlab for writing all the data processing and fitting programs used to extract peak parameters. One dimensional scattering data were obtained by azimuthal average of the two dimensional scattering images. Scattering data were normalized by the incident beam intensity, which was continuously monitored. These normalized data were corrected by subtracting the background and solvent scattering. The scattering data reached a constant intensity at high q due to the effects of thermal density fluctuations. After normalization and background subtraction this constant value was almost unchanged over the temperature range of 120-150 °C. Jeong et al[15] and Sakurai et al[27] treated this effect by subtracting the constant from the scattering intensity. We accounted for this constant term in the fitting procedure used to determine the peak parameters. Because of the very large data sets analyzed in this work, instead of determining these parameters by visually inspecting the data, we use an automated least squares peak fitting procedure using a Gaussian function to describe the peak shape. The scattering data in the primary peak region was fitted by adding two Gaussians to extract the peak intensity, position and widths of the two peaks in this region. The error in the parameters estimated by the fitting procedure is much smaller than the limiting resolution $\Delta q$ set by the finite pixel size. The peak position is fit to an accuracy of about 0.3% ( i.e. less than 0.001 nm$^{-1}$). The width is not equally well determined; it can have an error of about 3-5%. The extracted peak intensity and positions agree very well with those observed visually. We note that this is simply a phenomenological procedure to determine the time evolution of the peak parameters. To interpret the scattering in terms of the evolving structure of the block copolymer we calculate the scattering using a geometrical model with rippled lamellae for describing the transformation between LAM and HEX cylinder phases.

**Results and Discussion**

**Identification of Lamellar and HEX phases.** The phase diagram of SEBS in DEP has not been reported in the literature. Since the focus of our work is on understanding the kinetics we did not measure a detailed phase diagram, but instead examined SAXS from a few samples at different concentrations and temperatures to identify a sample which exhibits a LAM to HEX transition. Temperature ramp measurements on samples with concentrations of 20%, 25% and 30% showed only a HEX structure that transformed to disordered micelles on heating. SEBS in DBP at a concentration of 40% exhibited a HEX phase which transformed to a LAM phase on heating, and then was disordered at still higher temperatures. All measurements reported in this paper were made with this sample at 40% concentration. Static x-ray scattering measurements made at different temperatures are shown in Figure 1.

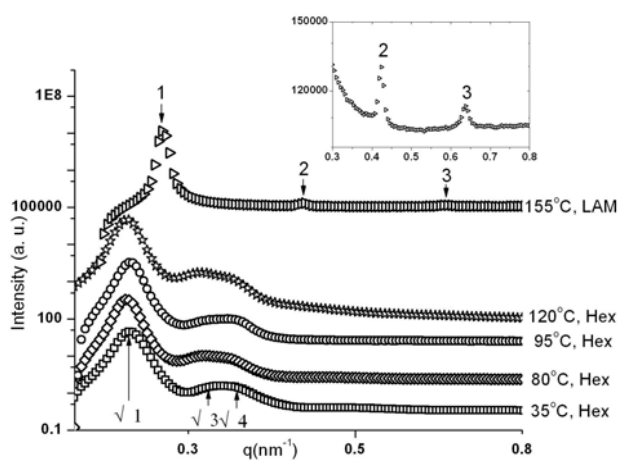

**Figure 1.** Static SAXS measurements at different temperatures, showing the HEX structure at lower temperatures (35 °C, 80 °C, 95 °C and 120 °C) and the LAM structure at higher temperature (155 °C). Scattering curves at the different temperatures are shifted along the intensity axis (by successive factors of 10 up to 120 °C and 1000 for 155 °C) to enable features to be seen clearly. The HEX peaks are labeled in the figure. The inset is a magnification of SAXS data at 155 °C in the q region of 0.3 nm$^{-1}$ - 0.8 nm$^{-1}$, to display the second and third order LAM peaks. The constant intensity at high q- due to thermal density fluctuations has not been subtracted in these plots because intensity is plotted on a log scale.

All samples were annealed for at least 2 hours before the SAXS measurement to ensure thermal equilibrium. The data was signal averaged for 2 minutes to obtain high signal-to-noise ratio. The relative positions of the Bragg peaks in Figure 1 clearly confirm that the sample is in a LAM state at 155 °C and HEX cylinder state at all of the lower temperatures that were measured. This observation of LAM phase at higher temperature is opposite to the result in most block copolymer melts, as mentioned in the Introduction. A melt of a very similar SEBS block copolymer (Shell G1652, 22.7 wt% PS) was examined by Jeong et al.[15] and showed the LAM phase at lower temperature than the HEX phase.

Clearly this difference is related to the differential solvent selectivity of DBP for the two component blocks. First, we note that SEBS in DBP forms reverse (or inverted) micelles with the majority component PEB in the cores of the segregated domains, which is opposite to the behavior in the melt with the minority component PS in the cores. Inverted micelles have been reported in SBS in styrene selective solvents by Zhang et al.[28] The phase diagram of a block copolymer in a selective solvent is determined by a complex interplay between the effects of solvent selectivity and block incompatibility. Huang and Lodge[29] showed that it is possible to have a LAM phase at higher temperatures than the HEX phase from a self-consistent mean field theory calculation of the phase diagram for AB diblock copolymer in a solvent selective for B blocks and poor to A blocks. They found that at a concentration $\phi = 30\%$, the LAM phase formed at higher temperature than the HEX when the fraction of the solvent incompatible A block $f_A = 0.7$, using $\chi_{AS} = 0.8$ and $\chi_{BS} = 0.4$. Here $\chi$ denotes the usual Flory Huggins parameter between polymer A or B and solvent S. (See Figure 10 of ref. 29). For the SEBS triblock used in our study, $\phi$ = 40% and the fraction of the PEB block which is in poor solvent condition is 0.72. In order to compare the phase behavior of this sample with the Huang and Lodge calculations, we estimate the $\chi$ parameters for the solvent interaction using the relation between $\chi$ and solubility parameters[30] and obtain[31] $\chi_{PEB-DBP} \approx 0.62$ and $\chi_{PS-DBP} \approx 0.36$. For the polymer-polymer interaction parameter we use the approach of ref. 28 and approximate with the PS-PB system, $\chi_{PS-PB} N$ =106 at 120 °C.[32] Assuming that a symmetric ABA triblock can be compared to an AB/2 diblock, we compare our observation with Huang and Lodge's calculated result that LAM is at a higher temperature than HEX for $f_A$ = 0.7, $\chi_{AS}$ = 0.8 and $\chi_{BS}$ = 0.4, $\phi$ = 0.3 and $\chi N$ =50. The transition for the 40% SEBS in DBP sample occurs at somewhat lower value of $\chi_{AS}$ and $\chi_{BS}$ than those in the calculation, which is not surprising given the increased concentration and the other simplifying approximations. It is also worth noting that solvent selectivity itself depends on temperature. While $\chi_{PS-DBP}$ changes by a negligible amount (from 0.358 to 0.357) as temperature increases from 120 to 150 °C, $\chi_{PEB-DBP}$ *decreases* from 0.627 to 0.607 as temperature *increases*, i.e. preferential solvent selectivity of DBP *decreases* with *increasing* temperature. The decrease of solvent selectivity with increasing temperature was also seen by Lodge et al[33] for poly (styrene-*b*-isoprene) in DBP.

**Determination of LAM to HEX transition temperature**. To determine the HEX → LAM transition temperature, a T-ramp was performed from 95 °C to 155 °C at a heating rate of 1 °C/min. The SAXS data acquired during the T-ramp shown in Figure 2(a) clearly show a structural change. Figure 2(b) shows selected frames in the vicinity of the transition temperature. The OOT temperature was determined to be in the vicinity of 135 °C, as the √3 HEX peak began to diminish around that temperature, indicating the initiation of the LAM-HEX transformation. Due to the finite rate of heating we note that these temperatures are *overestimates* of the equilibrium OOT.

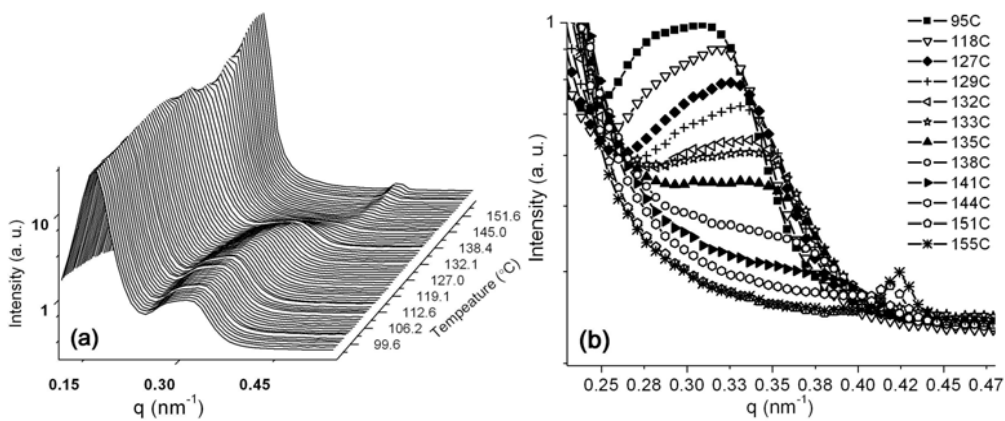

**Figure 2.** (a) Time-resolved SAXS data collected during a temperature ramp from 95 °C to 155 °C at a rate of 1 °C/min. (b) Several frames in the q range of 0.23-0.48 nm$^{-1}$ at different temperatures selected from the data shown in (a) to identify the transition temperature. The OOT temperature was located around 135 °C according to this temperature ramp experiment.

**Temperature Jump Measurements of Kinetics of LAM to HEX transition.** To follow the kinetics of the transition from the LAM to the HEX phase, a time-resolved SAXS experiment was done with a rapid T-jump from 155 °C → 120 °C going through the OOT temperature. The temperature reached 120 °C in less than 3 minutes, at a quenching rate about 0.2 °C/s. The scattering data is plotted in Figure 3, showing the LAM to HEX phase transformation.

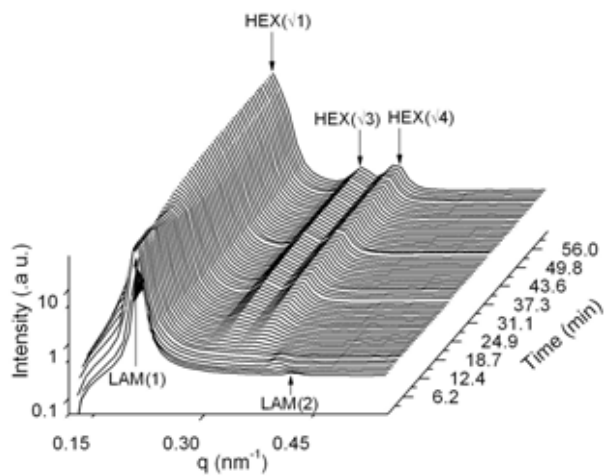

**Figure 3.** The time-evolution of SAXS intensity following a T-jump from 155 °C to 120 °C. The triblock copolymer SEBS transformed from LAM phase into HEX phase within 20 minutes.

Within a few minutes after the T-jump, structural changes are clearly visible, with three new Bragg peaks emerging. Two of them were in the positions before the second LAM peak, and one appeared right before the primary LAM peak. These new peaks correspond to the developing HEX structure, according to the relative positions of the Bragg peaks. The HEX peaks intensity increases while the LAM peaks diminish. After 20 minutes the HEX structure dominates in the scattering, as can be seen in Figure 4(a). The primary HEX peak is very close to the primary LAM peak and they are not well resolved. This can be seen via the broadening of the primary peak region. The initial transformation stage is depicted in Figure 4(b), showing the details of the transformation process.

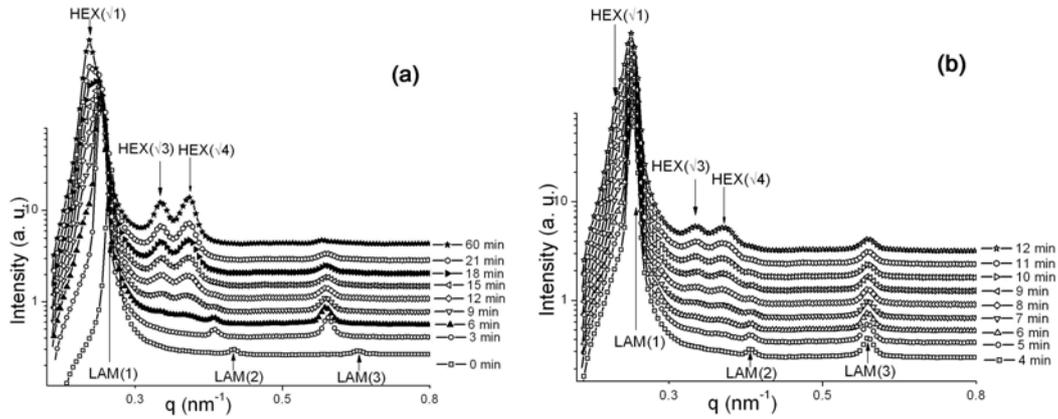

**Figure 4.** (a) Single frames at different time to show the complete LAM → HEX transformation. (b) Single frames from the earliest part of the data to show the details of the initial stages of LAM → HEX transformation.

**Analysis of T-jump SAXS data.** To study the kinetics of the transformation from the LAM phase to the HEX phase, we first need to quantitatively characterize the observed time evolution of the SAXS data. For this purpose, we used a simple Gaussian peak shape model to fit the time-resolved SAXS data in order to extract the peak intensity, position and width (full width at half maximum) of the scattering peaks. The data in the vicinity of the primary peak was fit to a sum of two Gaussians as described in the Methods Section. The fitted peak intensity and position agree very well with those observed by directly looking at the plots of the data. The time evolution of these parameters is shown in Figure 5(a)-(c). The apparent discrete change around 20 minutes in position and width is an artifact of the fitting method.[35]

The data show two stages during the transformation, as marked in Figure 5(b). Temperature equilibration occurs in stage (I) which shows only a LAM structure. The LAM transforms to the HEX in stage (II) followed by the growth of the HEX phase. Figure 5(b) shows that the position of the primary LAM peak decreases, i.e. lamellar spacing *increases* in stage I. In stage II the lamellar peak position is almost unchanged, but the HEX peak position decreases significantly during the initial part of stage II. During the early part of stage II there is a noticeable intensity from the LAM peak that decreases as the HEX intensity grows. We interpret this observation of HEX symmetry concurrent with a LAM structure in the early stages of the transformation as arising from scattering from correlated rippled lamellae, as discussed later. The changes of the intensity, position, and the width of the LAM and the HEX phase in stage (II) are very smooth, which strongly suggest that the transformation from the LAM to the HEX is a single step process of continuous ordering.[34] As shown in Figure 5(c), the width of the HEX phase decreases by a factor of 2-3, consistent with a factor of 2-3 growth in the size of the HEX domains. A log-log plot of the width versus time (plot not shown) shows that $W \sim t^{-\alpha}$ within the time interval from 5 to 30 minutes. The exponent $\alpha = 0.5$ implies that the average size of the ordered domains grows in time as $t^{1/2}$ (domain size is proportional to inverse of peak width).

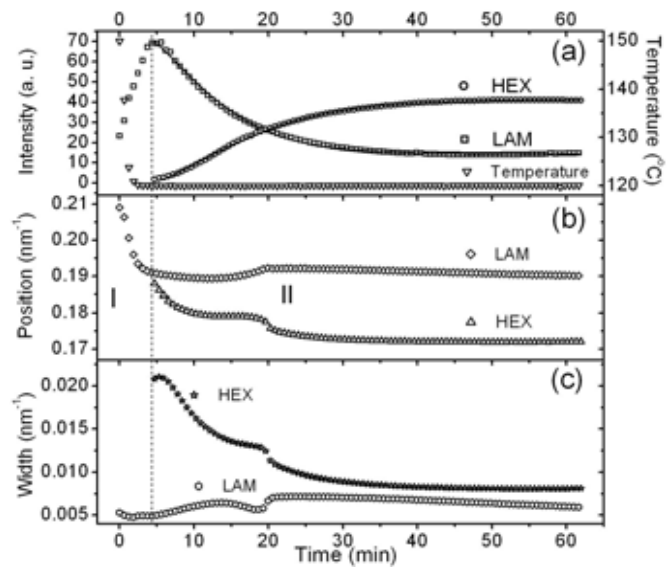

**Figure 5.** Time evolution of (a) the intensity of the LAM and the HEX primary peaks, and the temperature following the T-jump. The solid lines are the results of stretched-exponential fitting (discussed later) with exponent n = 1.3 for the LAM (squares) peak intensity and n = 1.35 for the HEX (circles) peak intensity. The LAM and the HEX primary peak (b) positions and (c) width as a function of time. The two stages during this transformation corresponding to (I) temperature equilibration and (II) transformation from the LAM to the HEX followed by the growth of the HEX are indicated. The apparent discrete change around 20 minutes in position and width is an artifact of the fitting method.[35]

**Transformation kinetics for different quench depths.** In contrast to the single step process that we have observed for the T-jump from 155 to 120 $^{o}$C, Hajduk et al.[10] observed a two-step transformation process in a melt of SEB using TEM and SAXS measurements. Most likely this difference is related to the quench depth of the kinetics process, as their measurements were for a very shallow quench (only a few degrees above the LAM-HEX phase boundary), whereas the measurement discussed above is for a very deep quench depth, $\Delta T = 35$ $^{o}$C. (We define the quench depth $\Delta T = T_i - T_f$, where $T_i$ and $T_f$ denote the initial and final temperature, respectively). Quench depth dependent differences of LAM to HEX transition kinetics have also been reported in melts of SEBS triblock copolymer.[15]

We examined the quench depth dependence of the kinetics by performing two other quenches, (i) a deeper quench from 150 C to 110 $^{o}$C ($\Delta T = 40$ $^{o}$C), and (ii) a shallower quench from 150 to 130 $^{\circ}$C ($\Delta T = 20$ $^{\circ}$C). In all three cases we observed the transformation from LAM to HEX, as expected because all the quenches cross the OOT around 135 $^{\circ}$C. We observed that the shallower the quench the slower was the transformation process. This is to be expected, because of the increased thermodynamic driving force for deeper quench depth. Furthermore $t_{ind}$, the induction time before which no appreciable transformation could be detected, *decreased with increasing quench depth.* For the shallowest quench ($\Delta T = 20$ $^{\circ}$C), which was just below the OOT, the induction time was about 25 minutes, much longer than that for the two deep quenches (4.5 minutes for $\Delta T = 35$ $^{\circ}$C and 3 minutes for $\Delta T = 40$ $^{\circ}$C). Such a monotonic decrease in $t_{ind}$ is commonly observed in a first order phase transition as the final quench temperature varies from near the coexistence temperature to the vicinity of the spinodal or the metastability limit. In a shallow quench, i.e. to temperature close to the OOT temperature and above the metastability limit phase separation proceeds via nucleation and growth, whereas in a deep quench, below the metastability limit, the system would be unstable and phase separate in a single step process, of continuous ordering, akin to the spinodal decomposition process. In both situations the transformation process would involve the growth of ripples. The distinction, as pointed by Matsen[36] in his studies of the HEX-BCC transformation (which also occurs by the growth of ripples), is that in the nucleation situation ripples develop at the growth front and these nucleated regions grow as the transformation proceeds, whereas in spinodal decomposition the ripples would grow spontaneoulsy in a correlated manner. Nucleation and growth is also characterized by an induction time as only nuclei larger than a critical grow and the system is metastable, whereas in spinodal decomposition the system is unstable and all fluctuations grow.

The time evolution of the SAXS intensity for the shallow jump ($\Delta T = 20\ ^oC$) is shown in Figure 6. As mentioned earlier we noted a long induction time, almost 25 minutes, and a very slow transformation process. The induction period corresponds to the formation of a metastable LAM phase. During the induction period the lamellae move further apart (all three LAM peaks shift to lower q); after this time the LAM spacing does not change as indicated by the constant peak position after 25 minutes. HEX peaks can be first identified at 25 minutes after which their intensity grows very slowly while the peak position remains fixed. During this growth stage the LAM intensity decreases a little while the position of the LAM peaks does not shift.

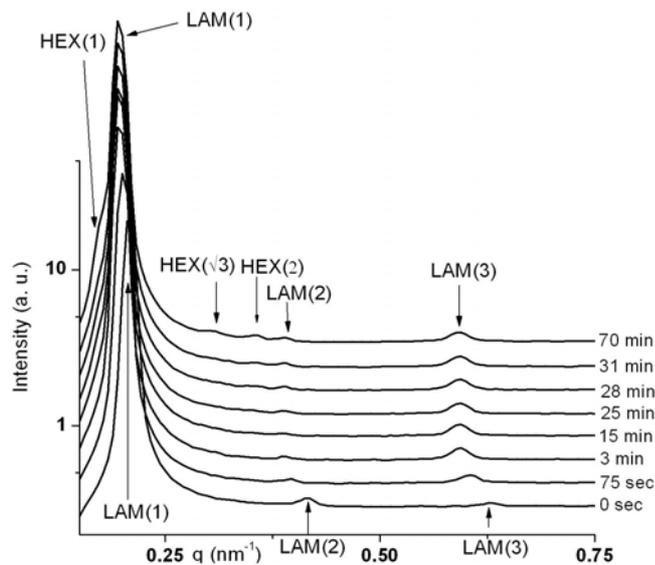

**Figure 6.** Single frames at different time to show the LAM → HEX transformation during a shallow temperature quench from 150 $^oC$ to 130 $^oC$. The HEX phase was first detectable at about 25 minutes after the temperature quench.

By the end of the run, only a very small fraction of the transformation is over, the two phases appear to coexist. It would be interesting to see if this phase coexistence persists for very long times. However, it was hard to perform very long run at the NSLS because of technical problems related to long-term beam stability and sample damage from prolonged exposure to a synchrotron beam.

Jeong et al[15] used time-resolved SAXS to study the LAM to HEX transition in a SEBS melt. They explained their observations in terms of a two-step nucleation and growth mechanism through a coexistence of HEX and LAM phases. We observe similar behavior for the shallow quench. To test the possibility of coexisting HEX and LAM phases for the deep quench we performed a linear weighted superposition of LAM and HEX scattering by adding the initial LAM phase scattering data at 155 °C with the scattering from the HEX phase at 120 °C (from the data shown in Figure 1) with weights *(1-f)* and *f*, where *f* denotes the fraction of the HEX phase in the structure. The superposition of scattering data with varying *f*, shown in Fig. 7, clearly shows that the peak position of the resulting scattering pattern does not shift with varying the fraction in the LAM phase. This is contrary to our observation that the HEX peak position shifts continuously while the LAM peak is more or less constant. The very small induction time observed in this deep quench data versus a shallow quench also indicate that this transformation is occurring in the vicinity of the metastability limit.

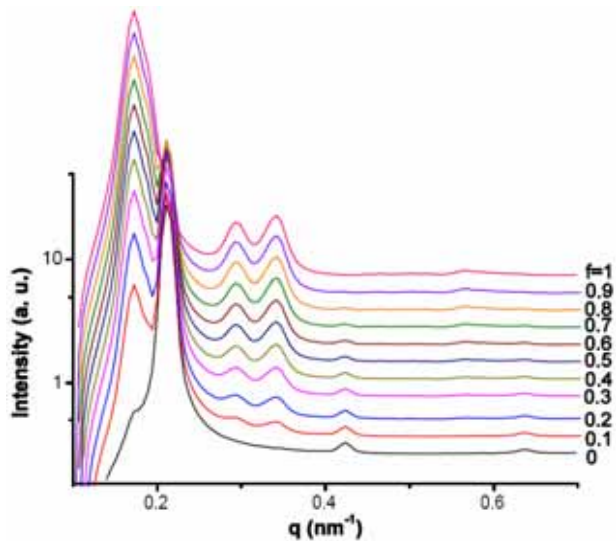

Figure 7. Linear weighted superposition of LAM and HEX scattering by adding the initial LAM phase scattering data at 155 °C with the scattering from the HEX phase at 120 °C with weights (1-$f$) and $f$, where $f$ denotes the fraction of the HEX phase in the superposition. The results show that the peak position of the resulting scattering pattern does not shift with varying the fraction in the two phases.

**Reversibilty of the LAM - HEX transition.** Following the temperature quench from 155 °C → 120 °C, we jumped the temperature back from 120 °C → 155 °C and maintained the sample at 155 °C while making time-resolved SAXS measurements. The reverse transformation from the HEX to the LAM phase was very slow, so the LAM phase did not develop fully even after 8 hours of annealing at 155 °C.

**Analysis of growth kinetics.** The time evolution of the peak intensity is related to the growth of the fraction of material that has transformed and is usually analyzed by the Avrami equation,[37]

$$I(t) - I(t_0) = (I(t_\infty) - I(t_0))(1 - e^{-k(t-t_0)^n}). \qquad (1)$$

As shown in Figure 5(a), the stretched exponential relaxation characteristic of the Avrami equation fits both the increasing intensity of HEX and decreasing intensity of LAM very well with exponent n = 1.3 for the LAM and n = 1.35 for the HEX (reduced $\chi^2$ = 0.4 and 0.13 for LAM and HEX respectively). The exponents are almost the same implying that this phase transition involves the formation of the HEX structure from the LAM structure followed by a coarsening process. The Avrami exponent we observed is quite different from that obtained Jeong et al[15], who found n < 1 for various temperature jumps for SEBS melts. Using the TDGL approach Qi and Wang[21] have obtained the time evolution of a global order parameter which is a measure of how the diblock melt transforms with time. Although a direct comparison of this order parameter with the scattering intensity was not attempted, because of the obvious difficulty of relating the global order parameter to SAXS intensity, it is worth noting that the temporal evolution of the Qi-Wang global order parameter shows the same trend as the stretched exponential fit of the Avrami equation to the intensity data.

**Geometrical Characteristics of the LAM and HEX Phases.** The spacing in the LAM phase is related to the position of the primary peak via, $d = 2\pi / q_1$, while the distance between nearest neighor cylinders in the HEX phase is given by $d = 4\pi / (\sqrt{3} q_1)$. From the peak positions for the LAM and HEX phases (see Fig. 1) we obtain that the spacing between adjacent lamellae is 30 *nm*, while the spacing between nearest neighbor cylinders in the HEX phase is 36 *nm*. This *increase* in spacing between cylinders in the SEBS in DBP is opposite to the *decrease* in spacing of the cylindrical phase observed in melts[9,15,18] and presumably reflects solvent selectivity effects.

The form factor for scattering from cylinders and lamellae can provide detailed information about the size of the scattering objects, as there are well-defined minima in these form factors whose positions depend on the size of the cylinders or lamellae.[38] The thickness L of the lamellae in the LAM phase and the radius R of the cylinders in the HEX phase can be calculated by locating $q_{min}$, the position of the first minimum after the primary peak using the relations $L = 2\pi/q_{min}$ and $R = 3.832/q_{min}$ which hold for the form factor of LAM and cylinder scattering.[38] These minima are hard to determine accurately, because they occur on the wings of scattering peaks. We obtain some estimates for these sizes using the long time averaged static SAXS data shown in Figure 1. A portion of this data is magnified and shown in Fig. 8. For the HEX phase a minimum can be observed around 0.46 nm$^{-1}$, while for the LAM data the minimum is harder to locate, but there may be one in the region 0.31 - 0.33 nm$^{-1}$. This suggests that the layer thickness $L \approx$ 19-20 *nm* and the cylinder diameter about 17-18 *nm* are of comparable magnitude.

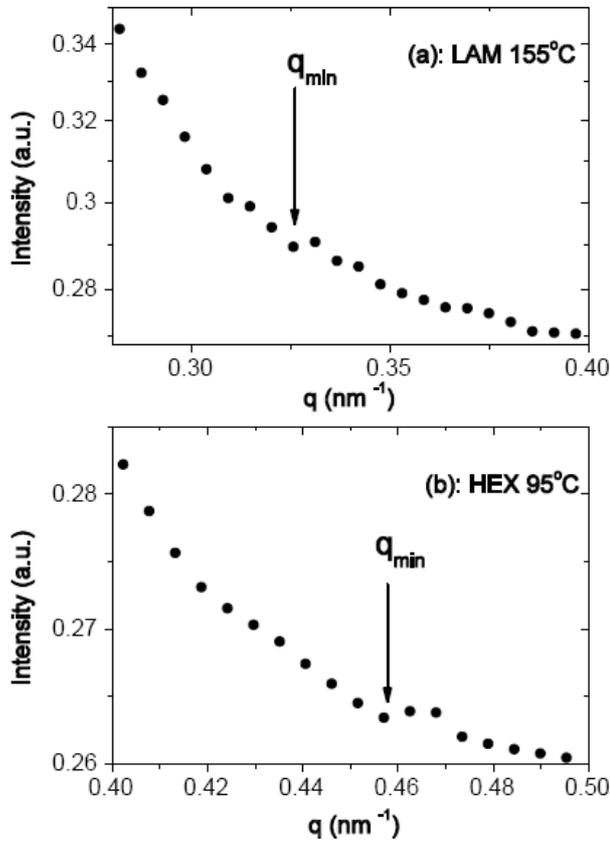

Figure 8. Estimation of $q_{min}$ from a portion of the static SAXS data of Fig. 1 shown here on a much higher magnification, (a) for the LAM phase at 155 °C, and (b) for HEX phase at 95°C. The estimated position of $q_{min}$ is indicated by the arrow.

The thickness of the PEB layer and the diameter of the cylindrical core can also be compared with $L_{PEB}$, the unperturbed end-to-end length of the middle PEB blocks. To obtain this length we use $L_{PEB} = a_{PEB}N_{PEB}^{0.5}$ treating the PEB chain which is in a poor solvent as Gaussian chain. We calculated the statistical segment length using the data for unperturbed chain dimensions of linear polymers from the Polymer Handbook[32a] and obtain $a_{PEB} = 0.68$ *nm* which gives $L_{PEB}$

=24.4 *nm*. This suggests that the PEB chains in the cylindrical core and the lamellae are compressed. Similarly, we compare the end-to-end length of the outside PS blocks in a good solvent with the void distance $D = d-2R = 18$ *nm* between the outer surface of the cores of two neighboring cylinders. For the PS chains we use the self avoiding walk chain formula $L_{PS} = a_{PS}N_{PS}^{0.6}$ since the PS blocks with $N_{PS}$ monomers are in good solvent. Using the statistical segment length $a_{PS} = 0.71$ *nm* we obtain $L_{PS} = 13.4$ *nm*.[32a] These values indicate that $L_{PS} < D < 2 L_{PS}$, implying that the PS chains of the coronas of neighboring cylindrical micelles interpenetrate, which is not surprising at a polymer concentration of 40%. We cannot obtain the length of the cylinders, because the minimum related to the length falls outside the SAXS q-range for the long cylinders that form in this block copolymer.

**Transformation Model for LAM to HEX**

As discussed above our experimental data for the deep jump suggest that adjacent lamellae are correlated when breaking into cylinders, which is a single step transformation process. The model of rippled lamellae transforming into cylinders has been extensively developed in previous studies.[8,20-22] We use this as a geometrical model to calculate the scattering from rippled lamellae and to analyze the transformation process from the LAM phase to the HEX phase. This transformation process is depicted in Figure 9. In the LAM phase, local density of SEBS in the layers varies on encountering fluctuations to the system, causing changes to the surface area. We assume the layers undulate like a ripple for simplicity, with an increasing amplitude *A* and fixed wavelength $\lambda$. Intuitively, the layer will break into separate cylinders when the undulation amplitude is big enough. When two adjacent layers break into cylinders, these cylinders must follow a kinetic path way such that the energy cost is at the minimum. Assuming all cylinders have the same dimension, the system gains the least free energy when cylinders of adjacent layers are totally out of phase, by minimizing the excluded volume interaction due to the PS coronas around the PEB cylinders. In contrast, uncorrelated cylinders formed by lamellae breaking randomly would require additional free energy to rearrange into a HEX lattice. In the model, the thickness of the layer in the LAM phase is $L$, the spacing between layers is $d$, the amplitude of the fluctuation is $a$, and the wavelength is $\lambda$.

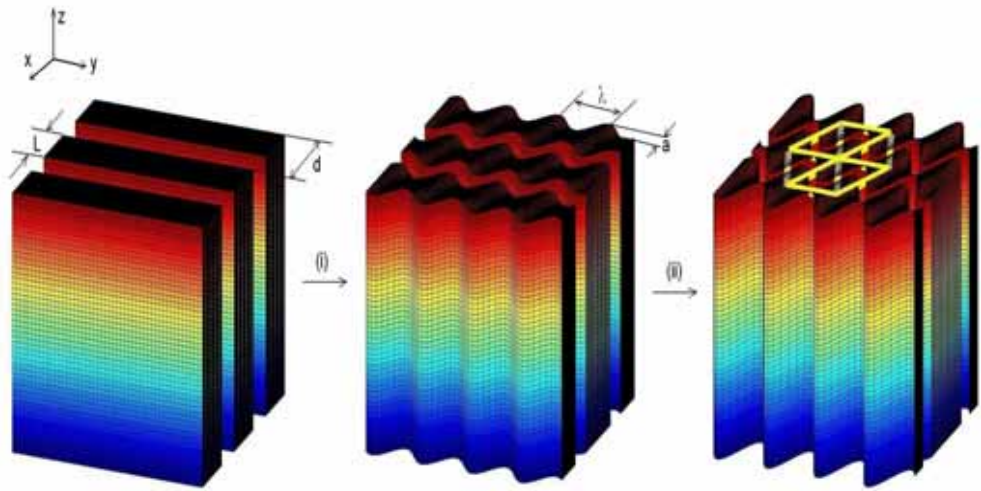

**Figure 9.** A schematic illustration of the transformation process from the LAM to the HEX phase. In the initial LAM phase, the thickness of the layer is *L*, and the spacing is *d*. In the earliest stages of the transformation labeled (i) layers in the LAM phase form ripple-like structures with wavelength $\lambda$ and amplitude *A*. At later stages (ii) the amplitude *A* is large enough to break the rippling structure into cylinders and form the HEX phase.

A commensurate HEX structure is formed if the spacing $d$ satisfies the following commensuration condition:

$$d = \frac{\sqrt{3}}{2}\lambda. \qquad (2)$$

In the incommensurate case, the model predicts an additional peak due to the rippling at $q'_1 = \frac{2\pi}{\lambda}\sqrt{1+\left(\frac{\lambda}{2d}\right)^2}$ close to the LAM peak at $q_1 = \frac{2\pi}{d}$. The two peaks coincide when the commensuration condition, Eq. (2) is satisfied. As discussed earlier, following a deep temperature jump we observed a new peak at a position just below the primary LAM peak. This corresponds to the predicted peak at $q'_1$ which eventually becomes the primary HEX peak.

We calculate the scattering intensity of the model structure and compare with our experimental data. The form factor $p(\vec{q})$ of a single lamella is,

$$\begin{aligned}p(\vec{q}) &= \int \exp(-i\vec{q}\bullet\vec{r})d\vec{r} \\ &= \frac{4\sin(q_y L/2)}{q_z q_y} \int_{-L/2}^{L/2} \exp(-iq_x x)\bullet\sin\left\{q_z\left[\frac{L_z}{2}+a\sin(2\pi\frac{x}{\lambda}+\phi)\right]\right\}dx\end{aligned} \qquad (3)$$

A unit cell in the transformation process consists with two adjacent lamellae; therefore the form factor $P(\vec{q})$ of the unit cell can be expressed as,

$$P(\vec{q}) = p(1) + p(2)\exp(-i\vec{q}\cdot\Delta\vec{r}_{12}) \qquad (4).$$

Where $p(1)$ and $p(2)$ are form factors of two adjacent lamellae, and $\Delta\vec{r}_{12}$ is the distance between two adjacent rippling cylinders of two adjacent lamellae. The structure factor $S(\vec{q})$ of the LAM lattice is,

$$S(\vec{q}) = 1 + \sum_{i=1}^{N} \exp(-i\vec{q} \cdot \Delta \vec{r}_i) \qquad (5).$$

The scattering intensity of an ordered domain can be written as,

$$I(\vec{q}) = |S(\vec{q}) \cdot P(\vec{q})|^2 \qquad (6).$$

Since the lamellae are not oriented, we average the scattering intensity over all directions,

$$I(q) = \frac{1}{4\pi} \int_0^{2\pi} d\Omega \int_0^{\pi} |S(\vec{q}) \cdot P(\vec{q})|^2 \sin\alpha \cdot d\alpha \qquad (7)$$

A representative set of scattering intensity calculated from the model is plotted in Figure 10, along with time-resolved SAXS data of SEBS 40 w/v% in DBP for comparison. We determined the wavelength $\lambda$ and the spacing $d$ from the data in stage II (see Fig. 5) using the predicted positions of the two peaks $q_1$ and $q_1'$ in the rippled LAM model. We found that the spacing d changed from 30.1 to 36.7 nm during the transformation period, while $\lambda$ changed from 41 to 42 nm. For simplicity we kept $\lambda$ fixed at 42 nm. The fluctuation amplitude $A$ grows as the transformation proceeds. A linear increase in the amplitude $A$ appears to agree quite well with the changes observed in SAXS data with time. When the fluctuation amplitude $A$ is 2 nm and the spacing $d$ = 31.4 nm, two peaks are clearly present in the first LAM primary peak region, indicating the formation of the rippled structure, as shown in Figure 10(a). This is almost the same as the SAXS data at 9 minutes shown in Figure 10(b). The primary peak position shifts to lower q value in both cases. When $A$ = 8-10 nm, the model calculation shows only one primary peak because the rippled lamellae are likely to break up into cylinders in the HEX phase. This agrees with the experimental observations at the late stages. In this case, the ratio d/λ is close to √3/2 in agreement with Equation 2. The calculation based on a model where adjacent lamellae are correlated during the transformation from the LAM phase to the HEX phase captures most of the features observed in the experiment. However, we have not attempted a detailed fit, or explored the effects of varying calculation parameters in detail.

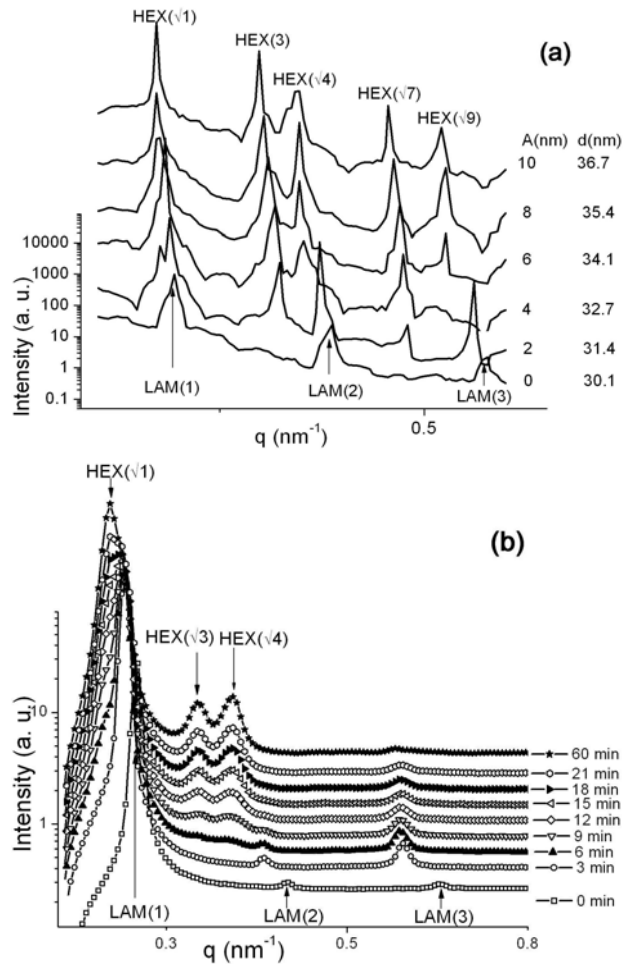

**Figure 10.** (a) The scattering intensity calculated from the model using 200 lamellae of thickness L=20 nm with wavelength $\lambda$=42 *nm*. The length of the lamellae along the direction of the wave was set to 50 $\lambda$. The fluctuation amplitude *A* was varied from 0 to 10 *nm* and the spacing *d* from 30.1 to 36.7 *nm* as indicated. (b) Time-resolved SAXS data of Figure 4(a) are reproduced here for comparison.

To further investigate the results of the model, in Figure 11 we plot the evolution of the peak intensities as the amplitude A grows and compare to SAXS data in the early stage from 4 minutes to 20 minutes of the LAM to the HEX phase transformation of Figure 5. The calculated intensity $I_{q_1}$ of the LAM peak at $q_1$ decreases exponentially, while the intensity $I_{q_1'}$ of the peak at $q_1'$ due to the ripples (which eventually form the HEX phase) increases exponentially, as shown in Figure 11(a). The change of the intensities with growing amplitude is similar to the time evolution of the peak intensities of the SAXS data as shown in Figure 11(c). The ratio of the shown in Figure 11(b) and 11(d). These similarities further confirm the validity of the geometric model we employed.

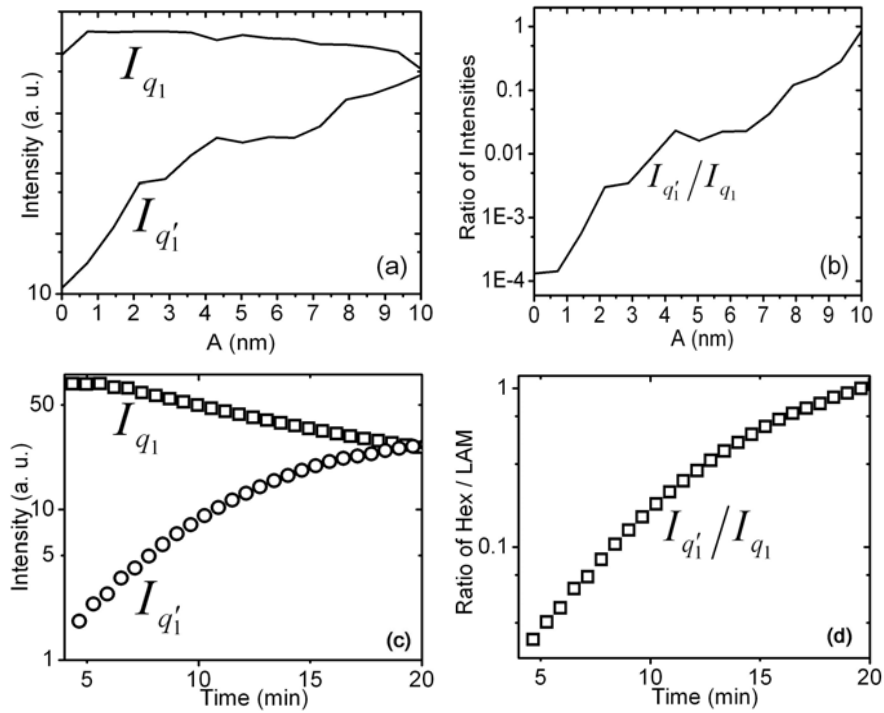

**Figure 11.** The intensities $I_{q_1}$ and $I_{q_1'}$ of the two peaks at $q_1$ and $q_1'$ calculated from the model and comparison with experimental data. All figures are in semi-log scale. (a) Calculated intensity of the $I_{q_1}$ and $I_{q_1'}$ for the model. (b) Ratio of the $I_{q_1'}$ and $I_{q_1}$ for the model. The ratio approaches 1 exponentially. (c) Peak intensities $I_{q_1}$ and $I_{q_1'}$, and (d) ratio of $I_{q_1'}$ to $I_{q_1}$ from the HEX and LAM intensities in the SAXS data.

The agreement between the calculation of the scattering from the geometric model and the experimental data for the deep quench supports the interpretation that the lamellae in the LAM phase transform to the HEX phase via a one-step rippling mechanism with adjacent layers totally out of phase, i.e. correlated undulation of the lamellar interfaces. For the shallow quench ripples nucleate at the cylinder growth front and the phase transformation occurs via two-step nucleation and growth process involving coexistence of HEX and LAM structures, similar to the observations of Jeong et al[15] in the SEBS melt. We note that in scattering measurements where the beam diameter is large compared to domain sizes it is not possible to distinguish between true phase coexistence and superposition of a HEX structure on a LAM structure. However, the continuous shift in the peak position for the deep quench data during the transformation stage (stage II) argues in favor of the HEX structure evolving by a continuous transformation of the LAM via a rippling mechanism for the deep quench.

**Conclusions**

We have measured the kinetics of the transformation from the LAM to the HEX phase for SEBS 40% w/v in DBP, a selective solvent for PS block using time-resolved SAXS. The solvent selectivity leads to inverted micelles containing the majority component (PEB) in the cores, and the occurrence of the HEX phase at a lower temperature than the LAM phase. The formation of inverted micelles and the inversion of the phase boundary is opposite to the behavior in the melt of SEBS. The SAXS data for a quench from 155 $^{\circ}$C in the LAM phase to 120 $^{\circ}$C, well below the OOT temperature, show that the primary peak position shifts continuously during the transformation indicating that the transition occurs via a one-step continuous ordering process. The spacing between the lamellae increases during the LAM to HEX transformation. We calculated the scattering using a geometric model of rippled layers with adjacent layers totally out of phase during the transformation. The agreement of the model with the experimental data lends further support to the continuous transformation mechanism from the LAM to the HEX for this deep quench. We also examine the quench depth dependence of the kinetics and found that the deeper the quench depth the faster is the process and the shorter the induction time for the HEX structure to appear. The data for a shallow quench from 150 $^{\circ}$C to 130 $^{\circ}$C, just slightly below the OOT of 135 $^{\circ}$C, show a very long induction time corresponding to the formation of a metastable LAM phase from which a HEX phase develops. For this shallow quench the data support a nucleation and growth mechanism during which coexisting HEX and LAM phases can be observed.

**Acknowledgements**

This research was carried out at Beamlines X27C and X10a of NSLS, Brookhaven National Laboratory which is supported by the U.S. Department of Energy, Division of Materials Sciences and Division of Chemical Sciences, under Contract No. DE-AC02-98CH10886. We thank Dr. Igor Sics, Dr. Lixia Rong and Mr. Steve Bennett for their support at beamlines X27C and X10a of NSLS. We acknowledge the support of Boston University's Scientific Computation and Visualization group for computational resources. We especially thank Dr. Huifen Nie for her help in the initial stages of this project. R.B. acknowledges the support of NSF Division of Materials Research Grant No. 0405628.**References**

(1) Leibler, L. *Macromolecules* 1980, 13, 1602.

(2) Bates, F. S.; Fredrickson, F. S. *Ann. Rev. Phys. Chem.* 1990, 41, 265.

(3) Matsen, M. W.; Bates, F. S. *Macromolecules* 1996, 29, 1091.

(4) Schultz, J. M. *Polymer Crystallization*, Oxford University Press, 2001.

(5) Matsen, M. W. *J. Phys.: Condensed Matter* 2002, 14, R21.

(6) Castelletto, V.; Hamley, I. W. *Current Opinion in Solid State and Materials Science* 2004, 8, 426.

(7) Almdal, K.; Koppi, K. A.; Bates, F. S.; Mortensen K. *Macromolecules* 1992, 25, 1743.

(8) Hamley, I. W.; Koppi, K. A.; Rosedale, J. H.; Bates, F. B.; Almdal, K.; Mortensen, K. *Macromolecules* 1993, 26, 5959.

(9) Sakurai, S.; Momii, T.; Taie, K.; Shibayama, M.; Nomura, S.; Hashimoto T. *Macromolecules* 1993, 26, 485.

(10) Hajduk, D. A.; Gruner, S. M.; Rangarajan, P.; Register, R. A.; Fetters, F. L.; Honeker, C.; Albalak, R. J.; Thomas, E. L. *Macromolecules* 1994, 27, 490.

(11) Foerster, S.; Khandpur, A. K.; Zhao, J.; Bates, F. S; Hamley, I. W.; Ryan, A. J.; Bras, W.
30

*Macromolecules* 1994, 27, 6922.

(12) Khandpur, A. K.; Forster, S.; Bates, F. S.; Hamley, I. W.; Ryan, A. J.; Bras, W.; Almdal, K.; Mortensen, K. *Macromolecules* 1995, 28, 8796.

(13) Radlinska, E. Z.; Gulik-Krzywicki, T.; Langevin, D.; Lafuma, F. *Langmuir* 1998, 14, 5070.

(14) Floudas, G.; Ulrich, R.; Wiesner, U.; Chu, B. *Europhys. Lett.* 2000, 50, 182.

(15) Jeong, U.; Lee, H. H.; Yang, H.; Kim, J. K.; Okamoto, S.; Aida, S.; Sakurai, S. *Macromolecules* 2003, 36, 1685.

(16) Ryu, D. Y.; Lee, D. H.; Jeong, U.; Yun, S.-H.; Park, S.; Kwon, K.; Sohn, B.-H.; Chang, T.; Kim, J. K.; Russell, T. P. *Macromolecules* 2004, 37, 3717.

(17) Loo, Y-L; Register, R. A.; Adamson, D. H.; Ryan, A. J. *Macromolecules* 2005, 38, 4947.

(18) Lai, C.; Loo,Y-L; Register, R. A.; Adamson, D. H. *Macromolecules* 2005, 38, 7098.

(19) Goveas, J. L.; Milner, S. T. *Macromolecules* 1997, 30, 2605.

(20) Qi, S.; Wang, Z.-G. *Phys. Rev. Lett.* 1996, 76, 1679; *Phys. Rev. E.* 1997, 55, 1682.

(21) Qi, S.; Wang, Z.-G. *Macromolecules*, 1997, 30, 4491; *Polymer* 1998, 39, 4439.

(22) Laradji, M.; Shi, A-C; Noolandi, J.; Desai, R. C. *Phys. Rev. Lett.* 1997, 78, 2577; *Macromolecules* 1997, 30, 3242.

(23) Wickham, R. A.; Shi, A.-C.; Wang, Z.-G., *J. Chem. Phys.* 2003, 118, 10293.

(24) Luo, K.; Yang, Y. *Polymer* 2004, 45, 6745.

(25) Nie, H.; Bansil, R.; Ludwig, K.; Steinhart, M.; Konak, C.; Bang, J. *Macromolecules* 2003, 36, 8097.

(26) Liu, Y.; Nie, H.; Bansil, R.; Bang, J.; Lodge, T.P. *Phys. Rev. E.* 2006, 73, 061803.
31

(31) To calculate the polymer-solvent Flory-Huggins $\chi$ parameter we use the Hildebrand solubility parameter approach (see Ref. 30) which gives

$$\chi_{polymer-solvent} = V(\delta_1 - \delta_2)^2 / RT + 0.34,$$

where V is the molar volume of the solvent, $\delta_1$ and $\delta_2$ are the solubility parameters of the polymer and the solvent. The solubility parameters of PS, PEB and DBP obtained from Ref. 30 are 18.6, 17.2, and 19.1 $[(J/m^3)^{1/2} \times 10^{-3}]$ respectively. Molar volume of DBP calculated from its molecular weight and density is $2.65 \times 10^{-4}$ m$^{-3}$.

(32) The polymer-polymer interaction parameter $\chi N$ for SEBS polymer is approximated using the approach of Ref. 28 for SBS with:

$$\chi N = \left(-900 + \frac{7.5 \times 10^5}{T}\right)\left(M_{w,A} v_A + M_{w,B} v_B\right),$$

where N is the degree of polymerization, $M_{w,A}$ and $M_{w,B}$ are weight average molecular weight of polymer A and B, $v_A$ and $v_B$ are the specific volume of polymer A and B. For SEBS weight average molecular weights of A block (PS) and B block (PEB) are 28 and 72 g/mol. The specific volumes of PS = 0.919 cm$^3$ and PEB = 1.114 cm$^3$ are obtained from ref. 32a below.

Kinetics of phase transition from lamellar to hexagonally packed cylinders for a triblock copolymer in a selective solvent

Yongsheng Liu, Minghai Li, Rama Bansil, Milos Steinhart

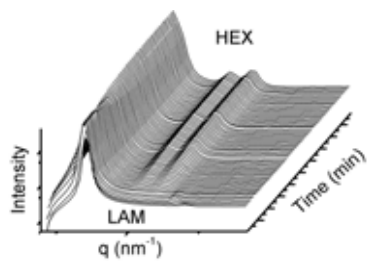